\documentclass[titlepage,aps,showpacs,prd,twocolumn,groupedaddress,superscriptaddress,floatfix]{revtex4-1}
\usepackage[utf8]{inputenc}
\usepackage{graphicx}
\usepackage{dcolumn}
\usepackage{bm}
\usepackage{amsmath}
\usepackage{amsfonts}
\usepackage{comment}
\usepackage{upgreek}
\usepackage{tabularx}
\usepackage{xcolor}

\usepackage[hidelinks]{hyperref}

\begin{document}

\title{Superdiffusive random laser}

\author{Federico Tommasi} 
\email{federico.tommasi@unifi.it}
\author{Lorenzo Fini} 
\author{Fabrizio Martelli} 
\author{Stefano Cavalieri} 
\affiliation{Dipartimento di Fisica e Astronomia, Universit$\grave{a}$
di Firenze, Via Giovanni Sansone 1 I-50019 Sesto Fiorentino, Italy.} 

\begin{abstract}
The peculiar characteristics of random laser emission have been studied in many different media, leading to a classification of the working regimes based on the statistics of spectral fluctuations.  Alongside such studies, the possibility to constrain light propagation by Lévy walks, i.e.\ with a `heavy-tailed' distribution of steps, has opened the opportunity to investigate the behavior of a superdiffusive optical gain medium, that can lead to a ``superdiffusive random laser.'' Here, we present a theoretical investigation, based on Monte Carlo simulations, on such a kind of medium, focusing on the widespread presence of fluctuation regimes, that, in contrast to a diffusive random laser, appears very hard to switch off by changing the gain and  scattering  strength. Hence, the superdiffusion appears as a condition that increases the value of the threshold energy and promotes the presence of fluctuations in the emission spectrum.

\end{abstract}

\maketitle
\section{Introduction}\label{cap:intro}
Among the stochastic processes, the random walk has been of  great importance in describing different phenomena in nature, such as the Brownian motion of particles in a fluid.  As an important case, the behavior of light propagation in a diffusive regime,  when the scattering mean free path is much grater than the radiation wavelength, can be efficiently described by a random walk. When the scattering interactions can be considered as independent and the first two moments of the path distribution are finite, the average square displacement $\langle x^2 \rangle$ increases linearly with time, leading to normal diffusion. On the contrary, anomalous diffusion can occur if the above conditions do not hold, such as in the case of scale invariant distributions of scattering events with diverging moments \cite{fract,pl1}.  In the general case:
\begin{equation}
\langle x^2 \rangle = Dt^\delta
\label{dif}
\end{equation}
where $D$ is the diffusion coefficient and $\delta\in (0,2]$  the scale parameter. Except for the cases of normal diffusion ($\delta=1$) and  pure ballistic propagation ($\delta=2$), the diffusion is anomalous:  $1<\delta<2$ for the \emph{superdiffusion} and $\delta<1$  for the \emph{subdiffusion}.

 Lévy walks and Lévy flights are particular cases of random walk that generalize and go beyond Brownian motion, leading to a superdiffusion dominated by the `heavy-tailed' probability distribution of the step lengths. These processes have been reported as characteristic of different motion patterns, such as animal foragin strategies \cite{PhysRevLett.88.097901,eco_lf} and human transport \cite{humantravel}. In properly engineered optical materials, called Lévy glasses, it  has been demonstrated that also  light can perform steps with a heavy-tailed distribution \cite{lf1,lf2,levyglass1}. 

Disordered materials with optical gain and normal diffusive properties, nowadays known as a source of random laser (RL) emission, have been extensively studied. 
RL is a particular optical source \cite{rl1,rl2} that has been the subject of growing interest since its first experimental realizations in the 1990's \cite{rl3,rl4,rl5,rl6}. The RL has an active material with scattering strong  enough to trap the radiation within, in such a way as to reach, without the presence of an optical cavity, the conditions of threshold, i.e., gain due to stimulated emission that overcomes losses. The achieved optical radiation shares some characteristics with conventional laser emission, such as the spectral narrowing, whereas the directionality and the spatial coherence is similar to the one produced by the spontaneous emission effect. Because of its high sensitivity to scattering, in recent years it has also been used for developing sensor devices  \cite{sensore,Tommasi:18}. 
\begin{figure}
\centerline{{\includegraphics[width=1.0\columnwidth]{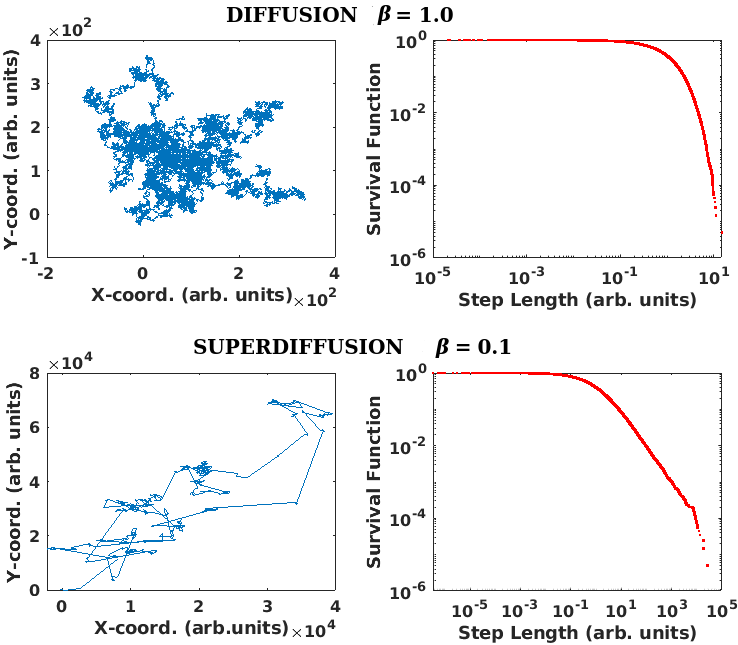}}}
\caption{(color online) Simulated paths of 10$^6$ steps in a two-dimensional medium where the starting site is placed in (x,y)=(0,0). In the first column, the simulated trajectories are shown with the same scattering characteristic length $\sigma$ = 1 (in arbitrary units) and different $\beta$: 1.0 (diffusion) in the first row and 0.1 (a case of superdiffusion) in the second one. In the second column, the corresponding survival functions, calculated by the empirical cumulative density function derived from the generated step lengths, are shown. }
\label{es_path}
\end{figure}

Great interest has been devoted to the theoretical and experimental investigation of the ``exotic'' and peculiar characteristics of random lasing; in particular, the emission behavior appears to be characterized by different regimes, with fluctuations that put RL in the field of the statistics of extreme events \cite{rl7,rl8,lepri,rl10,lepri2,Uppu:15,PhysRevA.98.053816,Sharma:06}.  Such an emission also manifests a replica symmetry breaking behavior \cite{rsb1,rsb2,rsb3,rsb4}. By means of the $\alpha$-index resulting from $\upalpha$-stable fit of the spectrum peaks, many studies have  led to a classification that is able to label the RL statistical regime \cite{rl9,rl11,rl12,nostro1,nostro2,PhysRevLett.114.183903,PhysRevA.90.025801}. Smooth and regular spectra are typical of the so-called Gaussian regimes ($\alpha\simeq 2$), whereas strong fluctuations are characteristic of the Lévy regime ($\alpha<2$).  In the latter, random spikes at random frequencies can manifest, leading to very different spectral output among different realizations, starting with the same initial conditions. 

In this paper we present a theoretical investigation about the emission behavior in a \emph{superdiffusive random laser} (SRL), i.e., an amplifying medium where the dispersion of the scattering points can be chosen with a ``heavy-tailed'' distribution. Hence, inside the active material, the light undergoes amplification by stimulated emission,  during a superdiffusion that can be tuned by a parameter that modifies the distribution of scattering interactions. In Sec.~\ref{cap:teo} the framework of the emission statistics of RL is summarized, as well as how  superdiffusion can be implemented in a Monte Carlo (MC) simulation. In Sec.~\ref{cap:mc} the details of the MC simulation are provided. In Sec.~\ref{cap:res} the results are presented and discussed, showing how the energy range affected by the fluctuation regime becomes larger in the case of SRL compared to the RL. In Sec.\ \ref{cap:for}, possible analogies with the field of ethology are discussed, while the conclusions are summarized in Sec.~\ref{cap:disc}.

\section{Theoretical background}\label{cap:teo}
\subsection{Previous model}
Random lasing is an optical phenomenon whos peculiar features are the due to interplay between the non linear mechanism of the stimulated emission and random scattering due to disorder. Several theoretical and experimental studies have been devoted to the statistical behavior of the intensity emission of the RL, leading to the detection of different statistical regimes in a way that is dependent on the pump energy \cite{rl9,lepri2,Uppu:15,rl10,rl11,rl12,nostro1,nostro2,PhysRevLett.114.183903,PhysRevA.90.025801,Sharma:06,raposo2015,lima2017,PhysRevA.98.053816}. For low pump energies the spontaneous emission is the ruling mechanism  and the output spectra are smooth and broad (first Gaussian regime). As the energy available in the medium increases, the contribution of stimulated emission becomes more important and the random laser can reach a threshold whose marker is the spectral narrowing. A further increase of initial energy leads to narrow and smooth spectra (second Gaussian regime) centered on the frequency with a more advantageous balance between gain and losses. In the energy range among the two Gaussian regimes, a fluctuation regime (L\'evy regime) can rise with in an  energy range that is dependent on the scattering strength.

 In a pure diffusive medium, when interference effects can be neglected, the output emission can be described as due to a very large number of extended modes that interact via the competition for the available gain.   Under this picture, the random spikes in the spectrum have a statistical origin, since they are produced by modes with a rare long lifetime  that is able to acquire a large amplification. In this case, their amplification is due to a non resonant feedback mechanism and these peaks called ``incoherent'' to distinguish them from the ``coherent'' peaks, due to random cavities established among a displacement of a large number of scatterers.
 
Considering an extended mode as a possible random path inside the medium, it can be demonstrated that the L\'evy regime is limited to a specific range of energies, with a crossover to a Gaussian regime for high and lower gain \cite{lepri}. Let $I(L)$ be the final intensity after a propagation with a total path $L$ inside a gain medium, 
\begin{equation}
I(L)\propto \exp[L/\ell_G],
\label{eq1}
\end{equation} 
where $\ell_G$ is the gain length. On the other hand, in the condition of normal diffusion, the probability density function (pdf) of $L$ is given by
\begin{equation}
f(L)=\frac{\exp[-L/\langle L \rangle]}{\langle L \rangle},
\label{eq2}
\end{equation} 
where $\langle L \rangle$ is the average length. Equations (\ref{eq1}) and Eq.~($\ref{eq2}$) lead to a power-law asymptotic behavior for the pdf of the intensity $I$,
\begin{equation}
g(I)\propto I^{-(1+\mu)}\qquad \mbox{where}\quad \mu=\frac{\ell_G}{\langle L \rangle}.
\label{eq3}
\end{equation}
The total intensity $I_T$ of the output emission is composed by the sum of single intensities, that are random variables distributed according Eq.~(\ref{eq3}), of a large number of modes. For $\mu\ge 2$, as a consequence of the central limit theorem, the fluctuations are Gaussian, while for $\mu<2$ the variance of the distribution diverges (L\'evy regime). 
At first glance, the increase of  initial energy stored in the medium would lead to a larger gain [$\ell_G\to 0$] and then to a power-law trend typical of a L\'evy regime. However, because of the competition between modes for the available gain, the possibility of amplification becomes in general dependent on time and spatial coordinates [$\ell_G=\ell_G(\vec{r},t)$]. Thus, as the initial energy in the system increases, the number of co-propagating modes increases too, leading to a rapid depletion of energy and then to a uniform saturation of the gain ($\ell_G\to\infty$). In such a condition of large energy, the extreme events in amplification are inhibited  and the fluctuations become Gaussian again. It is worth noting that high scattering promotes the gain competition between modes and then the fading of the energy range affected by the crossover of the L\'evy regime \cite{nostro1,nostro2,PhysRevA.98.053816}.  In a superdiffusive medium,  Eq.~(\ref{eq2}) no longer holds and in this paper we aim to study the consequences for the fluctuation regimes.
  As a summary, we report here the main reasons underlying the existence of the L\'evy regime: 
  
1) The scattering is not too strong, since strong scattering promotes the superposition of modes along their random paths and then the gain competition, averaging the amplification; and
2) the pump energy is not too high, because a large gain for a large number of modes causes a rapid depletion of energy stored in the medium, making rare long paths no longer able to achieve an anomalously large amplification. 

Hence, by changing the initial gain and the scattering properties of the medium, a variation and also a total inhibition of the energy range subjected to a Lévy regime can be achieved. 
It has also been reported that a simple parameter can be crucial to determine the emission statistics as the energy varies \cite{nostro1,nostro2},
\begin{equation}
\xi=\frac{\ell_s}{d}
\label{xi}
\end{equation}
where  $\ell_s$ is the scattering mean free path and $d$  the linear dimension of the active region of the medium. Hence, the energy range subjected to the Lévy regime can be controlled by the parameter $\xi$.  

The ``depth'' of such a fluctuation regime is described via an $\upalpha$-stable fit \cite{alfa1,alfa2,alfa3} of the peak value of a large number of spectra from random laser media with the same initial condition of scattering, total energy, and energy displacement geometry. A thin spectral window, centered on the frequency $\omega$, can be viewed as a total intensity $I_T(\omega)$ given by the sum of different modes  with the same frequency. The generalized central limit theorem states that the sum of a large number of  random variables with a pdf with a diverging variance approaches an $\upalpha$-stable distribution \cite{Lev,Gnedenko}. The main parameter of the $\upalpha$-stable pdf is the $\upalpha$-index, which describes the power-law asymptotic behavior:
\begin{equation}
P(I_T(\omega))\propto |I_T(\omega)|^{-(\alpha+1)}.
\label{asym}
\end{equation}
 Hence, for $\alpha<2$, a Lévy regime describes a more likely presence of fluctuations in the emission spectrum.
The Gaussian distribution is a special case of $\upalpha$-stable with $\alpha=2$ and it efficiently describes smooth spectra.   Since the convergence to  the Gaussian regime can be very slow, an arbitrary value of 1.8 can be chosen as a practical threshold value below which a Lévy regime is assigned. 

\subsection{Superdiffusive approach}
In the previous section, we discussed the theoretical background valid for a RL system where the Lambert-Beer (LB) law  is valid for describing both the amplification, i.e.\ by means of a negative absorption coefficient, and the scattering. In  SRL,   the LB law for the scattering is replaced with another suitable one. 

 Hence, let us focus on the intensity emission statistics of RL spectra to the scattering properties of the medium, where the distance between two consecutive scattering events is simply called a ``step''.
Without considering  amplification or absorption effects, in the case of  normal diffusion in a scattering medium, the step lengths $\ell$ are distributed according to the  LB law PDF,
\begin{equation}
p(\ell)=\frac{ e^{-\ell/\ell_s}  }{\ell_s}.
\label{lb}
\end{equation} 

In order to include the case of superdiffusion, the generalized Lambert-Beer (GLB) law  can be used \cite{mittag1,PhysRevE.77.021122},
\begin{equation}
p(\ell)=\sigma^{-\beta} \ell^{(\beta-1)}E_{\beta,\beta}\left(-\left(\ell/\sigma\right)^\beta\right) 
\label{lb2}
\end{equation}
where $\sigma$ is a scattering characteristic length, that decreases as the scattering of the medium becomes stronger, and $E_{\beta,\beta}$ is the two parameters Mittag-Leffler function \cite{mittag2,PhysRevE.77.021122}:
\begin{equation}
E_{\beta,\beta}(x)=\sum_{k=0}^\infty \frac{x^k}{\Gamma(\beta(k+1))}. 
\label{ml1}
\end{equation}
where $\beta\in (0,1]$ and $\Gamma$ is the Gamma function. For  $\beta=1$ (normal diffusion, that corresponds to the case of $\delta=1$ in Eq.~(\ref{dif})), Eq.\ (\ref{lb2}) becomes equivalent to Eq.\ (\ref{lb}) and $\sigma$ reduces to the  scattering mean free path $\ell_s$ ($\sigma=\ell_s$ for $\beta=1$). For $\beta\in(0,1)$ the PDF given by Eq.~(\ref{lb2}) has an `heavy-tail' with an asymptotic power-law trend $p(\ell)\sim\ell^{-\beta-1}$ \cite{mittag2} and represents a case of superdiffusion ($\delta\in(1,2)$ in Eq.~(\ref{dif})). The expression of Eq.~(\ref{lb2}), which was introduced to describe anomalous transport, is not the unique possible choice to describe superdiffusion. However, it has the advantage to including
 the  LB law for the case $\beta=1$ and not requiring any explicit cut for small $\ell$, as happens, for example, in the power law PDF case.  Moreover, for  the sake of clarity, we want to stress that in this paper we speak about ``diffusion'' and ``superdiffusion'' whenever scattering interactions can happen in the medium. Hence, also, weakly scattering media, where the hypotheses underneath the diffusion equation are not accomplished, are considered, for simplicity, as ``diffusive.''

In Fig.~\ref{es_path} different numerical simulations of paths composed by the same number of steps in a two-dimensional passive infinite medium are shown.  In the first row the $\beta$-parameter is 1 (diffusion), whereas in the second one it is 0.1 (a case of superdiffusion). In both cases the scattering characteristic length $\sigma$ is 1 (in arbitrary length units). In the case of superdiffusion, long steps are more probable, as it is shown in the second column of the figure, where the corresponding empirical survival function (defined as 1-ecdf, where ecdf is the empirical cumulative density function) of the generated steps are reported.   

This paper is devoted to the study of a random laser in the presence of superdiffusion. In particular, the emission properties are studied and compared to the case of normal diffusion in a medium with a homogeneous gain and a fixed $\sigma$, but a ``heavy tailed'' distribution of random scattering events.
Hence, here a SRL medium can be characterized by the following features:
(i) the gain properties (the total energy $\mathcal{E}$ and the linear size $d$ of the active region),
(ii) the scattering properties (the scattering characteristic length $\sigma$ and the Mittag-Leffler parameter $\beta$ of the distribution of steps lengths).

The generalization of  Eq.~(\ref{xi}) in the case of SRL  describes the scattering strength with respect to the size of the active region,
\begin{equation}
\xi=\frac{\sigma}{d}
\label{xi2}
\end{equation}
The emission behavior of a set of simulations with the same values of the parameters above is then characterized by the $\alpha$-index of the  statistical regime.

The amplification is considered as a property of the bulk material and as independent of the behavior of the scattering. Thinking about a practical case, one can  address, for example, a Lévy-glass uniformly infiltrated (bulk + glass spheres) by a gain material. 

\section{Monte Carlo simulations}\label{cap:mc}
The Monte Carlo simulations are based on the method used in Refs \cite{lepri,nostro1,nostro2,PhysRevA.98.053816}, with the medium simulated by a square lattice where an amount of excited atoms is distributed in the cells. 
The extended modes, i.e.\ possible random paths, are simulated by random walkers, with an individual energy $n$ and a frequency $\omega$. Each walker originates in a cell by a spontaneous emission event,  propagates at constant velocity through the adiacent cells, changes in direction by scattering events, and is amplified by stimulated emission. The random walkers are processed in parallel to introduce the gain competition among modes. The initial energy is provided to the medium at time $t=0$  and then released out  by the walkers once they exit from the lattice. 
The side $l=c\,dt$ of the cell, where $c$ is the walker speed and $dt$ the temporal step of the simulation, 
is called  as ``one cell'' for simplicity and it is the unit of arbitrary length used in this work. 

The initial energy, corresponding to an amount of excited atoms, is introduced in a circular region of $d$ cell diameters in a lattice of side $L$ cells.  Within this active zone the gain is homogeneous as long as the energy is not depleted by the propagating modes. The emission spectrum is simulated by frequency windows of 1001 (-500 to 500) arbitrary channels (arb.\ chs.), where the channel 0 corresponds to the center of resonance of the simulated atomic transition. 

\begin{figure}
\centerline{{\includegraphics[width=1.0\columnwidth]{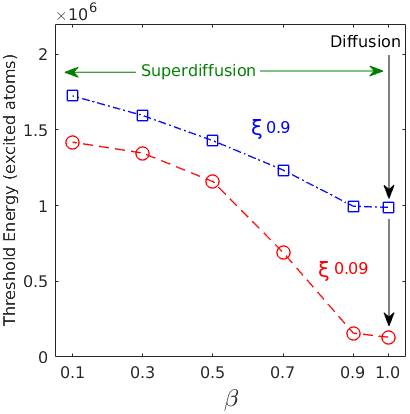}}}
\caption{(color online) Threshold energy as a function of $\beta$ for the medium with $d$ = 80 cells. The blue squares are the threshold values for the scattering characteristic length $\sigma$ = 70 cells ($\xi\simeq 0.9$) and the red circles are the ones for $\sigma$ = 7 cells ($\xi\simeq 0.09$). }
\label{fig_th}
\end{figure}

\begin{figure}
\centerline{{\includegraphics[width=1.0\columnwidth]{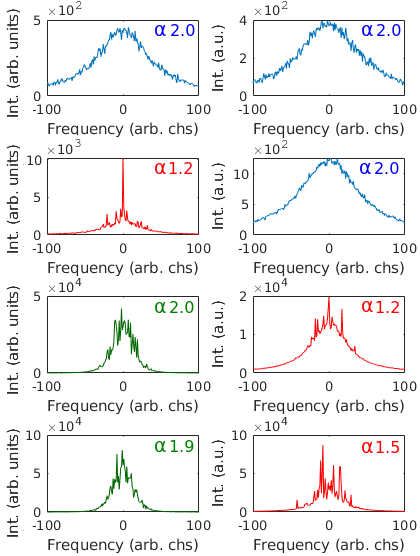}}}
\caption{(color online) Typical spectra for two media with the same $\sigma$ = 7 cells and $d$ = 80 cells at different initial energies in diffusion ($\beta$ = 1.0, on the left) and in a case of superdiffusion ($\beta$ = 0.3, on the right). The statistical regime found upon the analysis of all spectra at the same parameters is shown in the inset on top right-hand side of each case (arb.\ chs.\ = arbitrary channels).}
\label{fig_spettri}
\end{figure}

The simulation consists in a loop of three steps for each time interval $dt$, untill the total time reaches 5$\tau$, where $\tau=1/\gamma_0$ is the spontaneous emission lifetime. 
The first one simulates the \emph{spontaneous emission}:
In a cell with a number $N_j$ of excited atoms, a new random walker can be created with a probability $\gamma_0 N_jdt$, where $\gamma_0$ is the spontaneous emission rate in unit of $dt^{-1}$. Once born, a random walker carries $n=1$ of arbitrary units of energy and a frequency drawn from a Cauchy distribution of random numbers centered on the channel 0, to simulate the emission linewidth with a full width half maximum (FWHM) of $w$ of arb.\ chs. The initial direction of the trajectory is uniformly random.

The second step is the \emph{diffusion} of the random walkers within the lattice. 
At this point, a remark concerning the implementation of the pdf given by the Eq.~(\ref{lb2}) has to be considered. A generalized radiative transfer equation, which takes into account any suitable pdf for random paths, has been previously derived by Larsen \emph{et al.} \cite{LARSEN2011619}. By using this approach photon migration in conditions of anomalous diffusion can be simulated by a classical MC procedure where the classical $p(l)$ derived from the LB law [Eq.~(\ref{lb})] is replaced by a more general $p(l)$, such as Eq.~(\ref{lb2}).
Such a method allows us to describe a physical system with a Monte Carlo, circumventing the use of non-local operators.

 Hence, at the first instant of the life of a random walker and at each scattering event, a step length $\ell$ (in cells units) is drawn following Eq.~(\ref{lb2}), with a fixed $\sigma$ (in cells units) and $0<\beta\leq 1$ (fixed for each set of simulations). Such a numerical procedure is obtained by the inversion of the cumulative distribution associated with  Eq.~(\ref{lb2}), once given $u,v\in(0,1)$ two uniformely distributed random numbers  \cite{PhysRevE.77.021122}, 
\begin{equation}
\ell(u,v)=-\sigma\ln(u)\left[  \frac{\sin\left[ \beta\pi(1-v)  \right]  }{\sin[\beta\pi v]     }      \right]^{1/\beta} .
\label{ml2}
\end{equation}
For $\beta=1$ (normal diffusion), Eq.~(\ref{ml2}) reduces to the usual inversion formula for the exponential distribution that rules the Lambert-Beer law [Eq.\ \ref{lb}]:
\begin{equation}
\ell(u)=-\ell_s\ln(u)\qquad u\in(0,1). 
\label{ml3}
\end{equation}
where $\sigma$ becomes $\ell_s$ and $u\in(0,1)$ is a uniformly distributed random number.

After each time step $dt$, $\ell$ is decreased by 1 unit of the cell and the walker propagates to an adjacent cell according to its trajectory. When $\ell$ reaches 0, a scattering event occurs and a new direction and a new $\ell$ is drawn by Eq.\ (\ref{ml2}). If a walker reaches the sample boundary, it becomes part of the output emission and its energy $n$ and its frequency are recorded. 

The third step of the simulation is the \emph{stimulated emission}: The energy $n_i$ carried by the $i$th random walker and the local population $N_j$ of the cell of the lattice are deterministically updated with the following rules:
\begin{equation}
n_i\to[1+\gamma(\omega_i)N_jdt]n_i,
\label{stim1}
\end{equation}
\begin{equation}
N_j\to[1-\gamma(\omega_i)n_idt]N_j,
\label{stim2}
\end{equation}
where the stimulated emission coefficient $\gamma$ depends on the random walker's frequency $\omega_i$:
\begin{equation}
\gamma(\omega_i)=\frac{\gamma_0}{1+(\omega_i/w)^2}.
\label{stim3}
\end{equation}

\section{Simulation Results}\label{cap:res}
The common parameters for all the simulations are $\gamma_0=10^{-4}$ $dt^{-1}$ (hence the spontaneous emission lifetime $\tau$ is $10^{4}$ $dt$), $w=100$ arb.\ chs of frequency and a total time of $5\tau$ in $dt$ temporal units. The other parameters $\sigma$, $d$, initial energy and $\beta$ are characteristic of a specific set of simulations consisting in 200 runs with identical starting conditions.
\begin{figure}
\centerline{{\includegraphics[width=1.0\columnwidth]{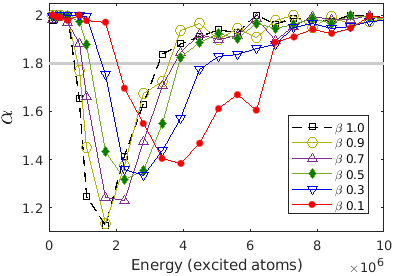}}}
\caption{(color online) $\alpha$-index as a function of energy for different paths distribution. $\sigma$ and $d$ are 70 and 80 cells, respectively ($\xi\simeq 0.9$).  The light gray line indicates the arbitrary value below which the L\'evy regime is established.}
\label{alfaZ}
\end{figure}
\begin{figure}
\centerline{{\includegraphics[width=1.0\columnwidth]{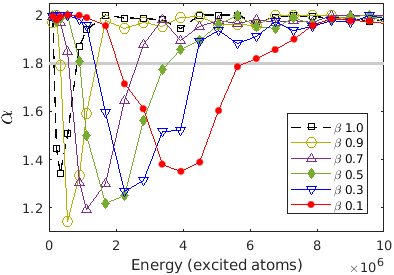}}}
\caption{(color online) $\alpha$-index as a function of energy for different paths distribution. $\sigma$ and $d$ are 7 and 80 cells, respectively ($\xi\simeq 0.09$). The light gray line indicates the arbitrary value below which the L\'evy regime is established.} 
\label{alfaY}
\end{figure}
\begin{figure}
\centerline{{\includegraphics[width=1.0\columnwidth]{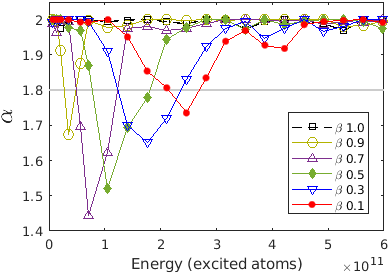}}}
\caption{(color online) $\alpha$-index as a function of energy for different paths distribution. $\sigma$ and $d$ are 5 and 250 cells, respectively ($\xi = 0.02$). The light gray line indicates the arbitrary value below which the L\'evy regime is established.}
\label{alfaW}
\end{figure}
\begin{figure}
\centerline{{\includegraphics[width=0.9\columnwidth]{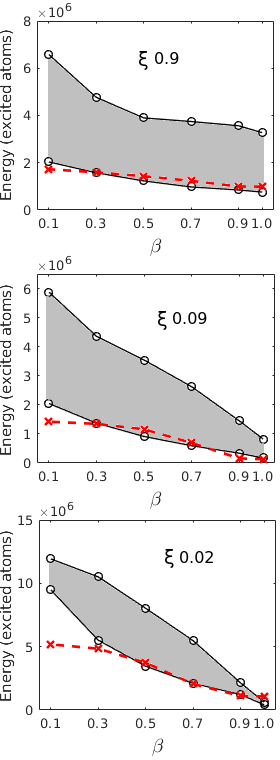}}}
\caption{(color online) Extension of the Lévy regime in the case of the medium for different value of $\xi$.  The gray area is the energy range where the L\'evy regime occurs. The threshold value (red dashed curve and crosses) is also shown. }
\label{alfa_wide_W}
\end{figure}

The parameter $\beta$ of  Eq.~(\ref{lb2}) is peculiar in this work and characterizes 
the transition from a diffusive regime ($\beta=1$) to a superdiffusive one ($\beta<1$). The chosen $\beta$ parameters for the simulated SRL are 0.1, 0.3, 0.5, 0.7 and 0.9, besides the value 1 for the ``conventional''  RL. 

As mentioned in the previous section, we use the ratio $\xi=\sigma/d$ to characterize the strength of the scattering with respect to the dimension of the active zone of the medium. 
The values of $\sigma$ are chosen to be $7$ cells ($\xi\simeq 0.09$) and $70$ cells ($\xi\simeq 0.9$) in the media with $d$ = 80 cells with a total dimension of 150x150 cells. Another set of simulations, with $\sigma$ = 5 cells and $d$ = 250 cells in a medium of 250x250 cells  has been performed to achieve a smaller value for $\xi$ (0.02). 

In Fig.\ \ref{fig_th} the threshold value for the media with $d$ = 80 cells is shown as a function of $\beta$. Since a random laser does not have a clear threshold value as in the case of the standard laser \cite{PhysRevE.65.047601}, in order to estimate it, the linear fit at high energies of the peak of the average spectrum is prolonged to the 0-peak value, considering the intercept with the energy axis \cite{Tommasi:18}. In the diffusive cases ($\beta=1$), as expected, a lower $\sigma$ leads to a smaller threshold value, because the photons lifetime inside the medium becomes larger. As expected, the superdiffusion causes the threshold to grow, because the long paths quickly bring the photons out of the medium, reducing the action of the stimulated emission.

The superdiffusive cases confirm that a strong scattering (small $\sigma$) leads to a lower threshold value,  with a decreased sensitivity on $\sigma$ as the medium becomes more superdiffusive ($\beta\to 0$). 

In Fig.~\ref{fig_spettri}, typical spectra are shown for the two cases, diffusion ($\beta=1$) and superdiffusion ($\beta=0.3$), at four different initial energies. The medium has $\sigma$ = 7 and $d$ = 80 cells. In the diffusive case (left column) the behavior is the one that was just characterized in the previous works \cite{nostro1,nostro2}, with an initial sub-threshold regime of smooth spectra (first Gaussian), a Lévy regime that is triggered just above threshold and then a second Gaussian regime at higher energies, with narrow spectra with moderate fluctuations. In the superdiffusive case (right column), the increase of the threshold value leads to an enlargement of the energy range characterized by the sub-threshold First Gaussian regime. Also in this case, the onset of the Lévy regime appears to be linked to the threshold value. As an important difference, the Lévy regime is mantained also for high energy. The $\alpha$-index of the statistical regime that is assigned to these cases is reported in the inset of each spectrum.
 
The extensive characterization of the statistical regimes is shown in  Figs \ref{alfaZ}, \ref{alfaY} and \ref{alfaW} for different $\beta$.  A L\'evy regime is considered as $\alpha$ falls below the arbitrary value of $\alpha$ = 1.8. The diffusive case confirms that the ratio $\xi=\sigma/d$ is critical to determine the width of the energy range affected by the Lévy regime, that appears to vanish as $\xi\to 0$. The superdiffusion, besides the shifting of the beginning of the Lévy regime to higher values, causes a broadening of the energy range characterized by strong fluctuations, but also with higher values of $\alpha$.

For the sake of clarity, such a behavior is also reported in Fig.\ \ref{alfa_wide_W}, where the gray area is the energy range characterized by the Lévy regime and the threshold value is shown as well.  

The general behavior is that the width of the levy region is increasing with  $\beta\to 0$ and this region rises just above the emission threshold. For the medium with strong scattering  ($\xi\to 0$) the L\'evy regime extension presents  clearly different behavior between diffusion and superdiffusion. This extension decreases with $\beta$ until reaching a complete closure for $\beta=1$.  At a low scattering  strength (larger $\xi$) the broadening of the L\'evy region increases with a moderate dependence on $\beta$.

To summarize the main results of these simulations for RL vs SRL, it can be noted that the threshold is increased and the L\'evy regime is broader and always present. The explanation lies in the scattering law in superdiffusion, causing the radiation  to escape from the medium in a shorter time, and then on average, with a smaller amplification. Thus, the threshold energy increases compared to the RL case. At the same time, the depletion of the medium is harder to be reached and then the rare long paths undergo an anomalously large amplification due to the mode uncoupling for the gain competition.
Such a condition persists despite the presence of high gain and/or scattering, causing a broadening in energy of the L\'evy regime in SRL.  A further argument for the broadening of the L\'evy regime is discussed in the next section.

\section{Random ``Foraging'' Efficiency}\label{cap:for}
As reported in previous works \cite{rl7,rl8,lepri,nostro1,nostro2,rl9}, in random laser systems in the diffusive regime, the random fluctuations can be framed as extended modes with rare long paths that become uncoupled by the competition between modes for the available gain. Then, such a large gain can be reached for a so-called ``lucky photon'' not only for its long lifetime, but also if the locations where the energy is distributed have not been ``visited'' by other modes or the same mode itself. Indeed, in the case of a large amount of energy stored in the medium, the gain competition leads to a rapid depletion, making the large paths no longer efficient and then inhibiting the fluctuations (second Gaussian regime) \cite{nostro1,nostro2}.  Hence, one should ask not only how rare  the long  paths are r inside the medium, but also how efficient  are  such paths to collect energy.

In  Sec.~\ref{cap:res}, larger values of threshold and broader Lévy regimes are reported in the case of superdiffusion. Due the high probability  of the emergence of large steps between scattering events, in average the random walkers spend a shorter time in the medium, leading to  a smaller mean amplification. In this section we present a further analysis addressed to explain the broadening of the L\'evy region by a model that is strictly connected to other scientific contexts. In ethology, the hypothesis of a greater efficiency of the Lévy walks in comparison with  Brownian motion has been studied with  growing interest \cite{PhysRevLett.88.097901,FERREIRA20123234,viswanathan}.  A searcher with \emph{a priori}  no information about the displacement of target resources can perform different kinds of random path patterns that can maximize the supply efficiency. Lévy-walk-like path patterns have been found in animal foraging strategies \cite{albatross,REYNOLDS20121225,evol2,Sims11073} and their hypothetical greater efficiency has been analyzed and discussed \cite{PhysRevLett.88.097901,eco_lf,REYNOLDS200998,Humphries7169,robot,evol2,BERBERT201841}.

Hence, it can be suggested to be ``exotic'' analogy, according to which the extended modes in RL or SRL act as ``foragers''  of excited atoms. An intuitive sketch, about the possibility to evaluate the L\'evy walks as more efficient, can be provided by Fig.~\ref{es_path}: The long steps in the superdiffusive case guarantee an exploration of a larger part of the medium and, in particular, prevent the ``oversampling'' of the same location yet depleted of resources. 
\begin{figure}
\centerline{{\includegraphics[width=1.0\columnwidth]{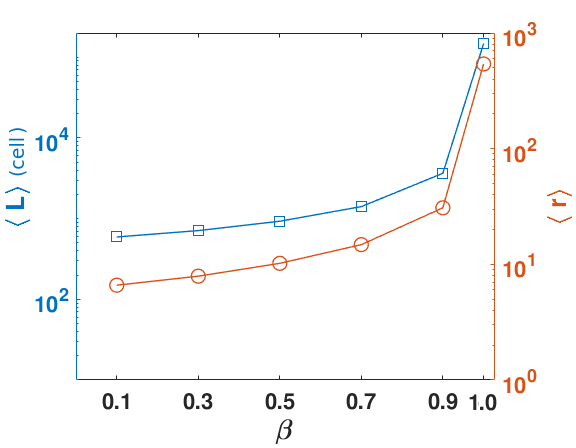}}}
\caption{(color online) The average of the total path length $\langle L \rangle$ (blue squares, left y-axis) and the average of the collected resource $\langle r \rangle$ (red circles, right y-axis). The diffusion guarantees a longer lifetime inside the medium and a larger number of collected resources.}
\label{for1}
\end{figure}
\begin{figure}
\centerline{{\includegraphics[width=1.0\columnwidth]{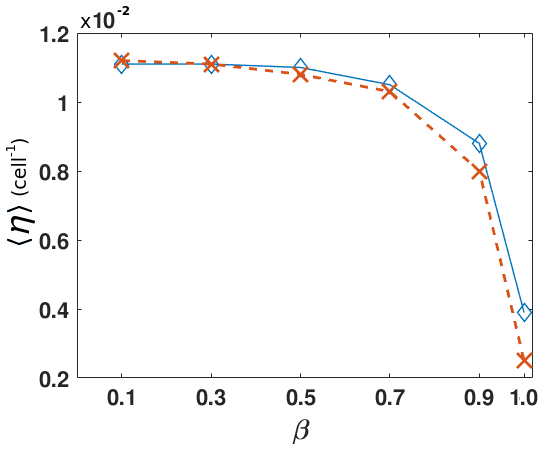}}}
\caption{(color online) The average specific harvesting  $\langle\eta\rangle$ as a function of $\beta$ calculated upon all paths (continuous blue  lines, diamonds) and upon the 0.01 percentile of the largest paths (dashed red lines, crosses). 
}
\label{for2}
\end{figure}

To support such a view, other sets of numerical simulations, based on a simpler model, have been performed. The scenario considers a double constraint on the random forager, since it performs a destructive search, in which  the reached resource becomes no longer available, until a boundary of the medium is reached and then the search ends with the destruction of the forager itself. 
In such a toy model, a forager propagates following a certain kind of random walk, starting from the center of a square medium lattice of 1000x1000 cells. The random step pdf, has a scattering characteristic length $\sigma$ of one cell and it is given by the Eq.~(\ref{lb2}).  Before the propagation of each walker, an amount of $10^4$ resources is randomly distributed in the medium, with a maximum of one resource per cell. Once the forager had entered a filled cell, the resource is collected and destroyed. The forager stops the search  when a boundary of the medium is reached and the collected resources and its total path length are registered. The total number of simulated foragers for each value of $\beta$ is $2\times 10^6$. We introduce here a simple parameter $\eta$ that measures the harvesting of resources for path unit (``specific harvesting''),
\begin{equation}
\eta=\frac{r}{L},
\label{eff}
\end{equation}
where $r$ is the number of collected resources and $L$ the total path length. 
In ethology $\eta$ is seen as the efficiency of the supply, that is crucial for the survival of the forager. 

In Fig.\ \ref{for1}, the average of the total path length $\langle L \rangle$ (blue squares, left y axis) and the average of the collected resources $\langle r \rangle$ (red circles, right y-axis) are shown. As expected, the diffusion regime ($\beta=1$) guarantees the longest forager lifetime and the largest average number of collected resources; in the  RL analogy, in diffusion regime the modes amplification allows a lower value of threshold energy, because the radiation spends more time within the active medium (as shown in Fig.~\ref{fig_th}) compared to the case of SRL.

In Fig.\ \ref{for2}, something different is underlined: The average (upon all the paths) specific harvesting $\langle \eta \rangle$   and the average specific harvesting $\langle \eta \rangle_{0.01\%}$ calculated only upon the largest paths ( the  0.01 percentile)   are reported as a function of $\beta$.
The $\eta$ appears to grow as the medium becomes more superdiffusive, as it has been reported  by ethology models and observations. In the context of the RL/SRL analogy, it is also important to note  that $\eta$, which is lower for the  largest paths in the diffusion regime, reach the same value calculated upon all the paths in superdiffusion regime (the two curves in Fig.~\ref{for2} tends to overlap as $\beta\to 0$).  The largest paths then  increase by their weight in the statistics, leading to a more likely presence  of extreme events in the evolution of the system, once the gain is added.  These characteristics  are in agreement with the broadening of the L\'evy region in the superdiffusive scattering condition.
  
\section{Conclusions}\label{cap:disc}
We have studied, by means of Monte Carlo simulations, the emission behavior expected for a random laser when the scattering properties are ruled by superdiffusion and compared the results to the usual case of normal diffusion. The main results show that, changing from diffusion to superdiffusion,
(1) The random laser threshold is increased, and (2)
 The L\'evy region is broadened and is always present.    
The first property can be explained by the fact that the radiation tends to escape from the medium in a short time, since large steps are more likely compared to the case of normal diffusion, and then achieves, on average, a smaller amplification. Moreover, since the depletion of the medium becomes inefficient in the case of superdiffusion, the rare long paths can more likely undergo an anomalously large amplification due to the mode uncoupling for the gain competition, despite the presence of high gain and/or scattering, causing a broadening in the energy of the L\'evy regime in SRL.  

In Sec.~\ref{cap:for}, we have reported a further argument for the main results of the paper in terms of ``specific harvesting''; 
 as reported in a similar way in ethology, the anomalously large path patterns can allow us to find resources in a more efficient way in the case of superdiffusion, avoiding oversampling of the same zones of the medium. 

As a prospective, such results should be verified in experimental samples of optical superdiffusive materials once gain is added.

 \section{Acknowledgments}\label{ringr}
Acknowledgments are due to  Francesca Tommasi and to Italia Comfidi Scarl for having provided the computers used for performing the MC simulations.

%

\end{document}